\documentclass[12pt]{article}
\usepackage{epsf}
\author{A.\,Oganesian\thanks{e-mail:armen@heron.itep.ru},
 A.\,Samsonov\thanks{e-mail:sams@heron.itep.ru}\\\\
\it Institute of Theoretical and Experimental Physics, \\
\it State Research Center, Moscow, Russia}
\bf
\title{Second moment of quark structure function \\
of the $\rho$-meson in QCD sum rules}

\begin{document}
\date{}
\maketitle
\newcommand{\qq}{\langle\overline{q}q\rangle^2}

\def\la{\mathrel{\mathpalette\fun <}}
\def\ga{\mathrel{\mathpalette\fun >}}
\def\fun#1#2{\lower3.6pt\vbox{\baselineskip0pt\lineskip.9pt
\ialign{$\mathsurround=0pt#1\hfil##\hfil$\crcr#2\crcr\sim\crcr}}}

$$\centerline{\hbox{Abstract}}$$
\rm
\indent The second moments of the quark structure functions of the 
longitudinally ${\cal{M}}_\rho^L$ and transversally ${\cal{M}}_\rho^T$ 
polarized $\rho$-mesons 
are calculated in the framework of 
QCD sum rules in external fields. The operators of dimension 4 
and 6 are taken into account. The results are ($\mu^2=1\, GeV^2$): 
${\cal{M}}_\rho^L=0.84\pm 0.08$, 
${\cal{M}}_\rho^T=0.5\pm 0.1$. 
The large difference between the values of ${\cal{M}}_\rho^L$ and 
${\cal{M}}_\rho^T$ indicates that the $\rho$-meson polarization strongly
influences the internal structure. In particular, in the case of 
the longitudinally polarized $\rho$-meson the gluon sea is found to be
strongly suppressed, it is less than $20\div 25\%$, whereas in usual hadron
case -- about $40\div 50\%$.
\\ \\ \\
\newpage
\section{Introduction}
$\,$\indent Quark structure functions determine 
quark distributions in the hadrons
and thus describe their dynamical properties. The structure functions of some
hadrons (pion or nucleon)  can be obtained from 
experimental data, while for other hadrons 
the necessary experimental information is
absent. That is why theoretical investigation of the structure functions 
is of great interest.\\
\indent The method of calculation of the structure functions
in the framework of QCD sum rules \cite{svz} was suggested in \cite{bel-iof}.
In \cite{iof-og-pion}, \cite{iof-og-rho} this method was generalized and 
the structure functions of the pion and $\rho$-meson were found. \\
\indent However, in the framework of the suggested technique the structure 
functions can not be calculated  at the values of the Bjorken
variable $x$, close to one or zero \cite{bel-iof}. Some additional
assumptions about structure functions behaviour in the region of large
$(0.7\la x<1)$ and small $(0<x\la 0.2)$ values of the Bjorken variable were 
used to describe the structure functions 
in the whole interval of $x$  \cite{iof-og-rho}.\\
\indent From this point of view the direct non-model calculation of 
second moments of the structure functions in QCD sum rules is very desirable.
The comparison of the values obtained with those
found in \cite{iof-og-rho} allows one to improve the structure functions
behaviour at small and large $x$. Also it is very interesting to verify 
the statement, proposed in \cite{iof-og-rho}, about large difference between
the structure functions of the longitudinally and 
transversally polarized $\rho$-meson.\\
\indent Furthermore, the values of the
second moment are of interest themselves, since they can be compared with 
those obtained from lattice simulations.
\\ \indent In the present paper we calculate the 
second moment of the quark structure function of the longitudinally and 
transversally polarized $\rho$-meson, using QCD sum rules in the external 
constant field. This technique was developed in \cite{iof-sm-magmom}. 
In this way the second moments of the structure
functions of the nucleon, pion, and kaon 
(\cite{kolesn}, \cite{bel-blok} and \cite{nishino}, where the quark masses
were taken into account) were calculated. \\
\indent In our paper $u$- and $d$-quarks are considered as massless,
$\alpha_s$-corrections are neglected. 
\section{Phenomenological part of the sum rule}
$\;$
\indent Let us consider the quark 
part of the symmetrized energy-momentum tensor:
$$
 T_{\mu\nu}={i\over 2}\sum_{\psi=u,d}
\overline{\psi}(\overrightarrow{D}_{\langle\{\mu} \gamma_{\nu\}}-
\overleftarrow{D}_{\{\mu} \gamma_{\nu\}\rangle}) \psi \,,\eqno(1)$$
where the curly brackets mean  
symmetrization over the indices, the angle brackets remind that the trace
must be removed,
$\overrightarrow{D_\mu} = \overrightarrow{\partial_\mu} -igA_\mu ^n t^n$, 
$\overleftarrow{D_\mu} = \overleftarrow{\partial_\mu} +igA_\mu ^n t^n$,
$t^n$ are Gell-Mann matrices. Hereafter we denote 
$\overrightarrow D_\mu$ simply as $D_\mu$.\\
\indent The second moment ${\cal{M}}_\rho$ of the $\rho$-meson 
structure function is known to be related to $T_{\mu\nu}$ in the 
following way (see, for example, \cite{nishino}):
$$
\langle\rho|T_{\mu\nu}|\rho\rangle
=2{\cal{M}}_\rho p_{\mu} p_{\nu}\;,
\eqno(2)$$
where $p_\mu$ is  the $\rho$-meson momentum, the traces are removed. \\
\indent In order to calculate ${\cal{M}}_\rho$ 
in the QCD sum rules technique one should
introduce a constant external field with a structure corresponded to
the operator $T_{\mu\nu}$ (1). 
That is why we consider the symmetric tensor field
with zero trace $S_{\mu\nu},\; S_{\mu\nu}=S_{\nu\mu},\;S_{\mu\mu}=0.$ 
Interaction with this field is determined by the additional term 
in the lagrangian: 
$$\Delta{\cal{L}}=-S_{\mu\nu}T_{\mu\nu}\,.$$
In this paper we consider the correlator of two vector 
currents with the 
$\rho$-meson quantum numbers in the external field $S_{\mu\nu}$:
$$\Pi_{\mu\nu}(p)=i\int d^4x\,e^{ipx}\langle T(j_\mu(x)j^+_\nu(0)\rangle_S\,,
\eqno(3) $$
where subscript S means presence of the external field. \\
\indent In this correlator the vector current has the form:
$j_\mu=\overline{u}\gamma_\mu d$, and its matrix element is 
$$\langle\rho^+\vert j_\mu \vert 0\rangle =(m_\rho^2/g_\rho)e_\mu\,,
\eqno(4)$$
where $m_\rho$ is the $\rho$-meson mass, $g_\rho$ is the $\rho$-$\gamma$ 
coupling constant, $g_\rho^2/(4\pi)=1.27$, and $e_\mu$ is the $\rho$-meson 
polarization vector. \\
\indent In the limit of weak external field we consider only linear in 
$S_{\mu\nu}$ terms in the correlator $\Pi_{\mu\nu}$ (3):
$$\Pi_{\mu\nu}=\Pi^0_{\mu\nu}+\Pi_{\mu\nu\rho\lambda}S_{\rho\lambda}\,.
\eqno(5)$$
\indent Let us construct sum rules for invariant functions at 
certain kinematical structures of $\Pi_{\mu\nu\rho\lambda}$ (5). 
According to the QCD sum rules technique, we represent 
these functions in two ways. On the one hand, we calculate them
at $p^2<0$ as the 
operator product expansion series. On the other hand, we saturate their 
dispersion relations  by the contributions of physical
states. Here we use the simplest model of physical spectrum, which contains 
the lowest resonance and continuum.
After equating of these representations the required sum rules
appear. \\
\indent Therefore, first of all one should choose the kinematical structures.
Let us suppose for a while that $\Pi_{\mu\nu\rho\lambda}$ is saturated
by the $\rho$-meson only. We denote such $\Pi_{\mu\nu\rho\lambda}$
as $\Pi^\rho_{\mu\nu\rho\lambda}$. One can extract an arbitrary 
invariant function $\Pi^i$ from $\Pi^\rho_{\mu\nu\rho\lambda}S_{\rho\lambda}$,
multiplying it by some projection operator $A^i_{\mu\nu}$:
$\Pi^i=\Pi^\rho_{\mu\nu\rho\lambda}S_{\rho\lambda}A^i_{\mu\nu}$. 
The dispersion relation for $\Pi^i$ has the form (we take into account (4)):
$$\displaylines{\Pi^i=\Pi^\rho_{\mu\nu\rho\lambda}S_{\rho\lambda}A^i_{\mu\nu}=
\sum\limits_{r,r^\prime}{\langle 0|j_\mu|\rho^r\rangle
\langle\rho^r|T_{\rho\lambda}|\rho^{r^\prime}\rangle\langle\rho^{r^\prime}|
j_\nu|0\rangle \over{(m_\rho^2-p^2)^2}}S_{\rho\lambda}A^i_{\mu\nu}=\hfill}$$
$$\displaylines{\hfill=
{m_\rho^4\over{g_\rho^2}}\sum\limits_{r,r^\prime}{e_\mu^r e_\tau^r
W_{\tau\sigma\lambda\rho} e_\sigma^{r^\prime} e_\nu^{r^\prime}\over
{(m_\rho^2-p^2)^2}}S_{\rho\lambda}A^i_{\mu\nu}\,,~~~~(6)}$$
where $\langle\rho^r|T_{\rho\lambda}|\rho^{r^\prime}\rangle S_{\rho\lambda}
=e_\tau^rW_{\tau\sigma\lambda\rho} 
e_\sigma^{r^\prime}S_{\rho\lambda}$, $W_{\tau\sigma\lambda\rho}$
is the amplitude of the
$\rho$-$\gamma$ scattering, $r,\,r^\prime$-- polarization indices (for the
transversal polarization $r,\,r^\prime=1,2$, for the longitudinal 
polarization $r,\,r^\prime$ have the only value). \\
\indent On the other hand (see (2)), 
$$W_{\mu\nu\rho\lambda}e_\mu^r e_\nu^r=
 2{\cal{M}}_\rho p_\rho p_\lambda \,,\eqno(7)$$ 
where the averaging over polarizations is assumed. \\
\indent Let us discuss the tensor structure of $W_{\mu\nu\rho\lambda}$. 
Starting from the most general form of $W_{\mu\nu\rho\lambda}$ and using
the equations $W_{\mu\nu\rho\lambda}p_\mu=W_{\mu\nu\rho\lambda}p_\nu=0$, 
which follow from the current conservation, we obtain:
$$\displaylines{W_{\mu\nu\rho\lambda}=
a\Big({1\over{p^2}}p_\mu p_\nu p_\rho p_\lambda-g_{\mu\rho}(p_\nu p_\lambda-
g_{\nu\lambda}p^2)-g_{\nu\lambda}p_\mu p_\rho\Big)+ \hfill}$$
$$+b_1(g_{\mu\nu}p_\rho p_\lambda-g_{\mu\rho}(p_\nu p_\lambda-
g_{\nu\lambda}p^2)-g_{\nu\lambda}p_\mu p_\rho)+\eqno(8)$$
$$\displaylines{\hfill+b_2\Big(g_{\mu\lambda}(p_\nu p_\rho-g_{\nu\rho}p^2)
-g_{\mu\rho}(p_\nu p_\lambda-
g_{\nu\lambda}p^2)+g_{\nu\rho}p_\mu p_\lambda 
-g_{\nu\lambda}p_\mu p_\rho\Big)
+b_3 g_{\lambda\rho}(p_\mu p_\nu -g_{\mu\nu}p^2)\,.}$$

\indent In order to find the invariants, related 
to the second moment, we
specify reference system. We work in the reference, where the $\rho$-meson 
momentum has the
form \\$p_\mu=(p_0,0,0,p_3)$ and the $\rho$-meson polarization vectors are: \\
$e^L_\mu=(e_0,0,0,e_3),\,e^L_\mu e^L_\mu=-1,\, e^L_\mu p_\mu=0$ 
for the longitudinal
polarization and \\$e^{T_1}_\mu=(0,1,0,0),\, e^{T_2}_\mu=(0,0,1,0)$ for 
the transversal polarization. \\ 
\indent We consider the certain components of the $\rho$-meson momentum
vector, for example, $\rho=3$ and $\lambda=0$. Using expression (8), 
we obtain for 
$W_{\mu\nu 30}e^r_\mu e^r_\nu$:
$$W_{\mu\nu 30}e^L_\mu e^L_\nu=ap_0p_3\,,$$
$${1\over 2}\sum_{i=1,2}W_{\mu\nu 30}e^{T_i}_\mu e^{T_i}_\nu=-b_1p_0p_3\,.$$
Considering equation (7) in the same reference and choosing 
$\rho=3$, $\lambda=0$, one can see that
the related to the second moment
invariants are: $a$ for the longitudinal and $-b_1$ for transversal 
$\rho$-meson polarization:
$$a=2{\cal{M}}_\rho^L\,,~~~~-b_1=2{\cal{M}}_\rho^T\,,\eqno(9)$$
where  ${\cal{M}}_\rho^L$ and ${\cal{M}}_\rho^T$ 
are the second moments of the quark structure 
functions of the longitudinally and transversally polarized $\rho$-meson.\\
\indent Substituting general form (8) into (6), we find
that $a$ is the only coefficient at the term 
$p_\mu p_\nu p_\rho p_\lambda S_{\rho\lambda}A^i_{\mu\nu}$ and $b_1$ is the 
only coefficient at the 
term $g_{\mu\nu}p_\rho p_\lambda S_{\rho\lambda}A^i_{\mu\nu}$. 
We come to conclusion that one should construct
the projection operator $A^i_{\mu\nu}$, which cuts the structure
$p_\mu p_\nu$ in the case of the longitudinal polarization
 and $g_{\mu\nu}$ in the case of the transversal polarization. 
However, practically instead of constructing the 
projection operators one can simply consider in $\Pi_{\mu\nu\rho\lambda}$ (5)  
the structure $p_\mu p_\nu p_\rho p_\lambda$ for the longitudinal
polarization and the structure $g_{\mu\nu}p_\rho p_\lambda$ 
for the transversal polarization.\\
\indent It is worth to note that such an analysis is not new, for example, in 
\cite{iof-og-rho} kinematical structures were chosen in the same manner. \\
\indent Now one can write down phenomenological representations of the 
invariants at the selected kinematical structures. 
Using formula (6) and taking into account also the contribution of the 
continuum, we obtain for the invariant at the
structure $p_\mu p_\nu p_\rho p_\lambda$:
$$\Pi^L(p^2)=\int ds{\rho_L(s)\over{(s-p^2)^2}}+...
\,,$$
$$\rho_L(s)=-{m_\rho^4\over{g_\rho^2}}{a\over s}\delta(s-m_\rho^2)+
f_L(s)\theta(s-s_\rho)\,,$$
and for the invariant function at the
structure $g_{\mu\nu}p_\rho p_\lambda$:
$$\Pi^T(p^2)=\int ds{\rho_T(s)\over{(s-p^2)^2}}+...\,,$$
$$\rho_T(s)=-{m_\rho^4\over{g_\rho^2}}b_1\delta(s-m_\rho^2)+
f_T(s)\theta(s-s_\rho)\,.$$
In these equations dots mean the contributions of non-diagonal 
transitions (for example, $\langle 0|j_\mu|\rho^\star\rangle\langle
\rho^\star|T_{\rho\lambda}S_{\rho\lambda}|\rho\rangle\langle\rho
|j_\nu|0\rangle$, where $\rho^\star$ is the excited state with 
the same quantum numbers as $\rho$), 
functions $f_L$ and $f_T$ represent continuum contribution 
and $s_\rho$ is the continuum threshold for the $\rho$-meson.\\
\indent Taking into account relations (9) and retaining only the terms, 
which do not vanish after Borel transformation, we obtain:
$$\Pi^L(p^2)=-{m_\rho^2\over{g_\rho^2}}
{2{\cal{M}}_\rho^L\over{(m_\rho^2-p^2)^2}}+
{\tilde{C}_L\over{m_\rho^2-p^2}}+
\int\limits_{s_\rho}^\infty ds{f_L(s)\over{(s-p^2)^2}}\,,\eqno(11)$$
$$\Pi^T(p^2)={m_\rho^4\over{g_\rho^2}}
{2{\cal{M}}_\rho^T\over{(m_\rho^2-p^2)^2}}+
{\tilde{C}_T\over{m_\rho^2-p^2}}+
\int\limits_{s_\rho}^\infty ds{f_T(s)\over{(s-p^2)^2}}\,.\eqno(12)$$
Here  ${\cal{M}}_\rho^L$ and ${\cal{M}}_\rho^T$ 
are the second moments of the quark structure 
functions of the longitudinally and transversally polarized $\rho$-meson,
$\tilde{C}_L,\;\tilde{C}_T$ appear due to non-diagonal transitions.
\\
\indent  In the next section functions $\Pi^L(p^2)$ and
$\Pi^T(p^2)$ in the left-hand sides of (11), (12)  will be calculated 
as the operator product expansion series. 
\section{Calculation of the vector current correlator}
$\,$ \indent Let us calculate in the correlator (3) the functions at 
the selected kinematical structures 
$p_\mu p_\nu p_\rho p_\lambda$ and $g_{\mu\nu}p_\rho p_\lambda$, $\Pi^L(p^2)$
and $\Pi^T(p^2)$ correspondingly, basing on the operator product expansion 
in QCD. \\
\indent First of all, we write down the quark propagator
in the external field $S_{\mu\nu}$ \cite{bel-blok} (fig.1):
$$\langle Tq_\alpha^a(x)\overline{q}_\beta^b(0)\rangle_S=
{i\delta^{ab}\over{2\pi^2x^4}}\biggl(\hat{x}+S_{\rho\lambda}
\bigg(x_\rho \gamma_\lambda
-{4x_\rho x_\lambda\over{x^2}}\hat{x}\bigg)\biggl)_{\alpha\beta}\,.$$
Here $\alpha,\,\beta$ are spinor indices, $a,b$ - color indices.\\
\indent An interaction with the soft gluon field modifies 
this expression\footnote{The 
proportional to $x_\eta x_\sigma x_\tau$ parts of these
expressions can be found in \cite{bel-blok}.} (we write down the proportional
to $S_{\rho\lambda}$ terms only):
$$\displaylines{\langle Tq_\alpha^a(x)\overline{q}_\beta^b(0)\rangle_{GS}=
\hfill}$$ 
$$\displaylines{\hfill
={gG_{\eta\sigma}^n(t^n)^{ab}S_{\rho\lambda}\over{8\pi^2x^2}}
\bigg({x_\rho x_\lambda \gamma_\eta \hat{x}\gamma_\sigma \over{x^2}}+
{x_\rho \over{4}}\big(\gamma_\lambda \gamma_\eta \gamma_\sigma-
\gamma_\sigma \gamma_\eta \gamma_\lambda \big)+{g_{\lambda\sigma}\over{2}}
\big( \gamma_\eta \hat{x} \gamma_\rho-\gamma_\rho \hat{x} \gamma_\eta \big)
 \bigg)_{\alpha\beta}}$$
(all propagators in fig.2),
$$\langle Tq_\alpha^a(x)\overline{q}_\beta^b(0)\rangle_{SGG}=
{g^2\langle G^2\rangle \delta^{ab}S_{\rho\lambda}\over{2^6 3^2 \pi^4}}
\int d^4z{z_\rho \over{(x-z)^4 z^2}}\Big((\hat{x}-\hat{z})
(\hat{z}\gamma_\lambda-z_\lambda)\Big)_{\alpha\beta}$$
(all propagators in fig.3).\\
\indent Let us begin with the case of the longitudinal polarization.\\
\indent We consider only the operators, which contribute to $\Pi^L(p^2)$ at 
the structure $p_\mu p_\nu p_\rho p_\lambda$, and retain only the terms, 
remained after Borel transformation. \\
\indent The contribution of
the loop diagrams (fig.4) to $\Pi^L(p^2)$ is equal to 
$$-{1\over{2\pi^2}}\int\limits_0^\infty{ds\over{(s-p^2)^2}}\,.$$
According to the quark-hadron duality, the continuum contribution in the
interval of $P^2=-p^2$ from $s_\rho$ to infinity is determined by the
bare loop in this interval. 
Therefore, function $f_L$ in (11) is constant: $f_L=-1/(2\pi^2)$.\\
\indent Diagrams, related to the operator $G_{\mu\nu}^n G_{\mu\nu}^n$,
are shown in fig.5a. The field induced  vacuum expectation value (fig.5b)
does not contribute to this kinematical structure. \\
\indent There are a number of vacuum expectation values of dimension 6
operators (see fig.6a). 
Some of them are not related to the external field. They are:
$$\langle \overline{q}_\alpha^a q_\beta^b D_\tau G_{\eta\sigma}^n\rangle=
{-g\qq\over{3^3 2^4}}(t^n)^{ba}(g_{\tau\sigma}\gamma_\eta-
g_{\tau\eta}\gamma_\sigma)_{\beta\alpha}\;,$$

$$\langle\overline{q}_\alpha^a (D_\tau q_\beta^b) G_{\eta\sigma}^n\rangle=
{-g\qq\over{3^3 2^5}}(t^n)^{ba}(g_{\tau\eta}\gamma_\sigma-
g_{\tau\sigma}\gamma_\eta-
i\epsilon_{\tau\eta\sigma\xi}\gamma_5\gamma_\xi)_{\beta\alpha}\;,$$

$$\langle\overline{q}_\alpha^a D_\sigma D_\eta D_\tau q_\beta^b\rangle=
{-ig^2\qq\over{3^5 2^4}}\delta^{ba}(g_{\eta\tau}\gamma_\sigma+
g_{\sigma\eta}\gamma_\tau-5g_{\sigma\tau}\gamma_\eta)_{\beta\alpha}\;.$$
The derivation of these expressions can be found in \cite{iof-sm-meson}.\\
\indent The only field induced vacuum expectation value, which contributes
to $\Pi^L(p^2)$, has the form (the last diagram in fig.6a):
$$\langle(D_{\{\sigma} D_\tau D_{\eta\}}  q_\alpha^a(0))
\overline{q}_\beta^b(0)\rangle_S= \delta^{ab}(c_1S_{\{\sigma\tau}
\gamma_{\eta\}}+
c_2g_{\{\sigma\tau}S_{\eta\}\chi}\gamma_\chi)_{\alpha\beta}\,.$$
The factors $c_1$ and $c_2$ are determined in Appendix. Only 
factor $c_1$ is involved in $\Pi^L(p^2)$. According to $(A12),\,(A13)$,
$$c_1=-3k(1+\epsilon_1)\,,~~~~k=-{ig^2\qq\over{3^52^4}}\,.$$
Here
$$\epsilon_1={ig\over{3k}}\Big(-d_1+{d_3\over 3}\Big)$$
represents the contribution of nonfactorizable vacuum expectation values
into $c_1$. In Appendix the set of unknown parameters $d_1,\,d_3,\,d_4$
is introduced and it is shown that all possible 
nonfactorizable field induced vacuum expectation values of dimension 6
operators can be expressed in terms of these three parameters. \\
\indent For $d_1,\,d_3,\,d_4$ the following relation takes place
$(A11)$:
$${4\over 3}d_1-d_3+d_4=-{g\qq\over{3^32^4}}\,.\eqno(13)$$
In assumption 
that the absolute value of each of these parameters is of the order of 
${g\qq\over{3^32^4}}$ (or less), 
one easily see that $|\epsilon_1|$ can reach the values
up to $1\div 2$, i.e. the accuracy of the estimation of $c_1$
is low -- about 100-200\%.\\
\indent Fortunately, the contribution of the related to  
$\langle(D_{\{\sigma} D_\tau D_{\eta\}}  q_\alpha^a(0))
\overline{q}_\beta^b(0)\rangle_S$ diagram to the total correction of the
dimension 6 operators is strongly numerically suppressed (the diagrams with
the hard gluon exchange are dominating). 
Quantitative analysis shows that large 
$|\epsilon_1|\approx 1\div 2$ results in 30\% uncertainty in the total
correction of the dimension 6 operators for
any more or less reasonable values of $d_1,\,d_3,\,d_4$. As we shall 
see, the operators of dimension 6 give less than 5\% of the value of the
second moment. Therefore, even 100\% uncertainty in $\epsilon_1$ does not
influence the accuracy of our results, and we can safely neglect it. \\ 
\indent Collecting all the terms,
we obtain for  $\Pi^L(p^2)$:
$$ \Pi^L(p^2)=-{1\over{2\pi^2}}\int\limits_0^\infty{ds\over{(s-p^2)^2}}+
{1\over{9}}{\langle(\alpha_s/\pi) G^2\rangle\over{p^6}}+
{56\over{27}}{g^2\qq\over{p^8}}\;.\eqno(14)$$
\indent Now we consider the case of the transversally polarized 
$\rho$-meson.\\
\indent Again, we take into account only the operators, which contribute
to the function $\Pi^T(p^2)$ at the structure $g_{\mu\nu}p_\rho p_\lambda$,
and omit the terms, vanished after Borel transformation. \\
\indent
The loop diagrams (fig.4) give to $\Pi^T(p^2)$ the following contribution:
$${1\over{2\pi^2}}\int\limits_0^\infty{sds\over{(s-p^2)^2}}\,.\eqno(15)$$
Therefore, function $f_T$ in (12) is: $f_T(s)=s/(2\pi^2)$.\\
\indent The diagrams with the gluon condensate (dimension 4) are shown 
in fig.5a. The contribution of these diagrams to $\Pi^T(p^2)$
has an infrared divergence in the chiral limit. To calculate it, we 
introduce nonzero mass $\mu_q$ for the quark, interacting with the 
external field $S_{\rho\lambda}$. We have:
$$-{2\over{9p^4}}\langle{\alpha_s\over{\pi}}
G^2\rangle \bigg(\ln{\Big(}{-p^2\over{\mu_q^2}}\Big)+
{1\over{3}}\bigg)\,.\eqno(16)$$
However, in the case of the transversal polarization one more 
vacuum expectation value of dimension 4 operator in the 
external field appears: 
$\langle(D_\sigma q_\alpha^a(0))\overline{q}_\beta^b(0)\rangle_S$ (fig.5b).
Its most general form is:
$$\langle(D_\sigma q_\alpha^a(0))\overline{q}_\beta^b(0)\rangle_S=
c_3S_{\sigma\tau}(\gamma_\tau)_{\alpha\beta} \delta^{ab}\,.\eqno(17)$$
Factor $c_3$ was found in the paper \cite{bel-blok-zf}. \\
\indent According to \cite{bel-blok-zf}, in order to obtain $c_3$,
one should construct a sum rule for the correlator 
$$\Pi_{\mu\nu\sigma\tau}(p)=i\int d^4x\,e^{ipx}\langle 
T(j_{\mu\nu}(x)j^+_{\sigma\tau}(0)\rangle\,,\eqno(18)$$
$$j_{\mu\nu}= {i\over 2}\sum_{\psi=u,d}
\overline{\psi}(\overrightarrow{D}_{\{\mu} \gamma_{\nu\}}-
\overleftarrow{D}_{\{\mu} \gamma_{\nu\}}) \psi\,.$$
The phenomenological part of the sum rule is 
saturated by the contributions of the $f_2(1270)$ meson and continuum. In
the operator product expansion part one should subtract not only perturbative 
contribution (15), but also the 
contribution of the operator $G_{\mu\nu}^n G_{\mu\nu}^n$, since it has been 
taken into account in (16). In such a way it was obtained \cite{bel-blok-zf}:
$$c_3={i\over{12}}\bigg({3s_f^2\over{320\pi^2}}-g_f^2m_f^4 -
{1\over{18}}\langle{\alpha_s\over{\pi}}
G^2\rangle \bigg(\ln{s_f\over{\mu_q^2}}-
{1\over{2}}\bigg)\bigg)\,.\eqno(19)$$
Here $m_f=1.27\,GeV$ is the $f_2$-meson mass, $s_f$ is the
continuum threshold for the $f_2$-meson and $g_f$ is the coupling constant 
of the $f_2$-meson with tensor current. Using (19), we can calculate the
diagrams in fig.5b. \\
\indent The total contribution of the dimension 4 operators 
to $\Pi^T(p^2)$ is equal to
$$-{1\over{p^4}}\bigg({3s_f^2\over{80\pi^2}}-4g_f^2m_f^4+
{2\over{9}}\langle{\alpha_s\over{\pi}}
G^2\rangle \bigg(\ln{\Big(}{-p^2\over{s_f}}\Big)+
{5\over{6}}\bigg)\bigg)\,.$$
It should be noted that this expression does not contain the quark mass
$\mu_q$, i.e. the infrared divergences cancel out. \\
\indent The operators of dimension 6 were discussed above (fig 6a). 
Besides them, new vacuum
expectation values of the operators of the same dimension appear (fig.6b):
$$\langle\overline{q}_\alpha^a G^n_{\mu\nu}D_\eta q_\beta^b\rangle_S\,,~~~~
\langle\overline{q}_\alpha^a\overleftarrow{D_\eta}G^n_{\mu\nu}
q_\beta^b\rangle_S\,,~~~~
\langle\overline{q}_\alpha^a(D_\eta G^n_{\mu\nu})q_\beta^b\rangle_S\,.
\eqno(20)$$
All of them are expressed in terms of parameters $d_1,\,d_3,\,d_4$
(see Appendix). \\
\indent The contribution of  
vacuum expectation values (20) is suppressed numerically even more than in 
the case of the longitudinal polarization (by the factor about 1/30).
Using the same estimation for
$d_1,\,d_3,\,d_4$ as in the previous case (see (13)),
 one can find that the 
uncertainty in the value of the total contribution of dimension 6
operators, which appears due to these parameters, is 
less than 15\%. This gives the uncertainty about few percents only in the 
final answer for the second moment. Therefore, one can neglect all the terms,
which contain $d_1,\,d_3,\,d_4$.\\
\indent It should be noted here that the field induced vacuum 
expectation value $\langle\overline q_\alpha^a\overline q_\beta^b q_\gamma^c 
q_\delta^d\rangle_S$ is zero (It can be shown with the help of the 
factorization hypothesis, see Appendix).\\
\indent As the result we obtain 
for the operator product expansion part of the sum
rule for the transversally polarized $\rho$-meson: 
$$\displaylines{\Pi^T(p^2)
={1\over{2\pi^2}}\int\limits_0^\infty{sds\over{(s-p^2)^2}}-\hfill}$$
$$\displaylines{\hfill-
{1\over{p^4}}\bigg({3s_f^2\over{80\pi^2}}-4g_f^2m_f^4+
{2\over{9}}\langle{\alpha_s\over{\pi}}
G^2\rangle \bigg(\ln\Big({-p^2\over{s_f}}\Big)+{5\over{6}}\bigg)\bigg)-
{140\over{81}}{g^2\qq\over{p^6}}\,.~~~~(21)}$$
\section{Results and discussion}
\indent To complete the calculations, we apply to the phenomenological 
(11), (12)
and operator product expansion (14), (21) parts of the sum rules   
the Borel transformation $\hat{B}(M^2)$,
$$\hat{B}(M^2)=\lim_{P^2,n \to \infty \atop{P^2/n=M^2}}
{(P^2)^{n+1}\over{n!}}\bigg(-{d\over{dP^2}}\bigg)^n\,,$$
where $P^2=-p^2>0$, and equate them.\\
We obtain for the longitudinal polarization:
$${g_\rho^2\over{m_\rho^2}}e^{m_\rho^2/M^2}
\biggl({M^2\over{4\pi^2}}\Big(1-e^{-s_\rho/M^2}\Big)+{1\over{36}}
{\langle(\alpha_s/\pi)G^2\rangle\over{M^2}}-{14\over{81}}{g^2\qq\over{M^4}}
\bigg)={\cal{M}}_\rho^L(M^2)+C_LM^2\,,\eqno(22)$$
and for the transversal polarization:
$$\displaylines{
{g_\rho^2\over{m_\rho^4}}e^{m_\rho^2/M^2}
\Bigg({M^4\over{4\pi^2}}\bigg(1-e^{-s_\rho/M^2}
\Big(1+{s_\rho\over{M^2}}\Big)\bigg)-\hfill}$$
$$\displaylines{\hfill-{3s_f^2\over{160\pi^2}}+2g_f^2m_f^4-
{1\over{9}}\langle{\alpha_s\over{\pi}}G^2\rangle
\bigg(\ln{M^2\over{s_f}}+{11\over{6}}-C_E\bigg)+
{35\over{81}}{g^2\qq\over{M^2}}
\Bigg)={\cal{M}}_\rho^T(M^2)+C_TM^2\,.~~(23)}$$
Here $C_E=0.577...$ is the Euler constant, $C_L,\,C_T$ appear due to 
the nondiagonal transitions. \\
\indent Generally speaking, the second moments in these equations 
are functions of the
squared Borel mass $M^2$ too. To analyze these sum rules we 
calculate the second moments at arbitrary fixed point $\mu$.
We take into account that the following linear combinations of
the second moments of the quark ${\cal{M}}_\rho$ and gluon 
${\cal{M}}_{\rho G}$ structure functions are renormcovariant \cite{bel-blok}:
$$\Big({16\over{25}}{\cal{M}}_\rho(M^2)
-{9\over{25}}{\cal{M}}_{\rho G}(M^2)\Big)
L^{50/81}=
{16\over{25}}{\cal{M}}_\rho(\mu^2)-{9\over{25}}{\cal{M}}_{\rho G}(\mu^2)\,
\eqno(24)$$
$${\cal{M}}_\rho(M^2)+{\cal{M}}_{\rho G}(M^2)=
{\cal{M}}_\rho(\mu^2)+{\cal{M}}_{\rho G}(\mu^2)=1.\eqno(25)$$
Here 
$$L={\ln(M^2/\Lambda_{QCD}^2)/{\ln(\mu^2/\Lambda_{QCD}^2)}}\,,$$
$\mu$ is the operator expansion normalization point.\\
\indent From (24), (25) one can express ${\cal{M}}_\rho(M^2)$ and
${\cal{M}}_{\rho G}(M^2)$ in terms of ${\cal{M}}_\rho(\mu^2)$ and
${\cal{M}}_{\rho G}(\mu^2)$ and substitute them into (22), (23). Excluding 
${\cal{M}}_{\rho G}(\mu^2)$ with the help of (25), we obtain the 
following sum rules. \\
\indent For the longitudinally polarized $\rho$-meson:
$$\displaylines{{9\over{25}}(1-L^{50/81})+
{g_\rho^2\over{m_\rho^2}}e^{m_\rho^2/M^2}L^{50/81}
\biggl({M^2\over{4\pi^2}}\Big(1-e^{-s_\rho/M^2}\Big)+{1\over{36}}
{\langle(\alpha_s/\pi)G^2\rangle\over{M^2}}-{14\over{81}}{g^2\qq\over{M^4}}
\bigg)=\hfill}$$
$$\displaylines{\hfill ={\cal{M}}_\rho^L(\mu^2)+C_LM^2\,.\;\;(26)}$$
For the transversally polarized $\rho$-meson:
$$\displaylines{{9\over{25}}(1-L^{50/81})+
{g_\rho^2\over{m_\rho^4}}e^{m_\rho^2/M^2}L^{50/81}
\Bigg({M^4\over{4\pi^2}}\bigg(1-e^{-s_\rho/M^2}
\Big(1+{s_\rho\over{M^2}}\Big)\bigg)-\hfill}$$
$$\displaylines{\hfill -{3s_f^2\over{160\pi^2}}+2g_f^2m_f^4-
{1\over{9}}\langle{\alpha_s\over{\pi}}G^2\rangle
\bigg(\ln{M^2\over{s_f}}+{11\over{6}}-C_E\bigg)
+{35\over{81}}{g^2\qq\over{M^2}}
\Bigg)=\;\;\;\;\;\;\;\;\;\;\;\;\;\;\;(27)}$$
$$\displaylines{\hfill ={\cal{M}}_\rho^T(\mu^2)+C_TM^2\,.\;\;\;\;\;\;\;\;\;}$$
\indent We use the following values of parameters: \\
$\mu=1\,GeV-$ the operator expansion normalization point, \\
$\Lambda_{QCD}=0.2\,GeV$,\\
$m_\rho=0.77\,GeV-$the $\rho$-meson mass, \\
$g_\rho^2/(4\pi)=1.27-$the $\rho$-$\gamma$ coupling constant, \\
$s_\rho=1.5\,GeV^2-$the continuum threshold for $\rho$-meson \cite{svz},\\
$\langle(\alpha_s/\pi)G^2\rangle=0.012\,GeV^4-$the gluon condensate  
\cite{svz}, \\
$g^2\qq=0.23\times 10^{-2}\,GeV^6-$the quark condensate \cite{svz}, \\
$m_f=1.27\,GeV-$the $f_2$-meson mass.\\
\indent The parameters $m_f$, $s_f$ and $g_f$ appear in sum rule for the
transversal polarization (27), because we consider 
the correlator of the tensor
currents (18) to calculate the nonfactorizible vacuum expectation value
$\langle(D_\sigma q_\alpha^a(0))\overline{q}_\beta^b(0)\rangle_S$. This 
correlator was considered in \cite{bel-blok-zf}, when the following sum rule
was obtained:
$${1\over{\pi M^2}}\int Im\Pi(s)e^{-s/M^2}ds={3\over{80\pi^2}}M^4-
{1\over{18}}\langle{\alpha_s\over{\pi}}G^2\rangle+
{1\over{9}}{g^2\qq\over{M^2}}\,,\eqno(28)$$
$$Im\Pi(s)=\pi m_f^6g_f^2\delta(s-m_f^2)+
{3s^2\over{160\pi}}\Theta(s-s_f)\,.$$
So, one can exclude $g_f$ from the sum rule (27) by direct substituting it
from (28). The continuum threshold for the $f_2$-meson $s_f$ can be found 
from the sum rule (28) for the $f_2$-meson mass. An analysis shows that 
$s_f\approx 2.5\div 2.7\,GeV^2$. For the further calculation we choose 
$s_f=2.6\,GeV^2$, the corresponding value of the coupling constant 
of the $f_2$-meson with tensor current is: $g_f=0.044$. \\
\indent One can extract the values of 
${\cal{M}}_\rho^L(\mu^2),\,{\cal{M}}_\rho^T(\mu^2)$ in two ways. 
We can approximate the left-hand sides of 
the equations (26), (27) by a straight line in the certain interval
of $M^2$ and find it ordinate
at zero Borel mass. The left-hand sides of (26), (27) as functions of $M^2$ 
are shown in fig.7. Another way consists in applying
of the operator $1-M^2(d/dM^2)$ to equations (26), (27). \\
\indent Both these ways lead to the same values:
$${\cal{M}}_\rho^L=0.84\,,~~~~{\cal{M}}_\rho^T=0.5\,. \eqno(29)$$
\indent The sum rule (26) for the longitudinal polarization gives the 
value of the second moment ${\cal{M}}_\rho^L$ with a very good accuracy. 
The left-hand side of (26) as a function of $M^2$ in the interval
$0.7\,GeV^2 \le M^2\le 1.1\,GeV^2$ 
is close to horizontal
straight line (the thin curve in fig.7). This means that the contribution
of nondiagonal terms is very small. \\
\indent The contributions of 
the dimension 4 and 6 operators give not more than 5\% of the total 
answer each. \\
\indent The accuracy in (29) is determined by the uncertainties 
in the values of the
quark and gluon condensates. The results (29) correspond to 
$g^2\qq=0.23\times 10^{-2}\,GeV^6$ \cite{svz}.
However, from the analysis of $\tau$-decay data \cite{iof-zyab}
the larger value of the quark condensate was obtained recently:
$$g^2\qq=(0.38\pm 0.12)\times 10^{-2}\,GeV^6\,.\eqno(30)$$
 These results 
 do not contradict one another within the errors, 
especially if one takes into account  
that (30) was obtained from the sum rule, in which the $\alpha_s$-corrections
were accounted, whereas in \cite{svz} they were not. Nevertheless, the 
difference between the average values of the quark condensate is about 50\%.\\
\indent The uncertainty in the value of the gluon condensate is 
also about 50\%. \\
\indent 
Taking into account all uncertainties discussed, one can find that the 
accuracy of the value of the second moment in the case of the longitudinal
polarization is not worse than $5\div 10\%$. \\
\indent A variation of the 
continuum threshold for the $\rho$-meson $s_\rho$ 
in the reasonable limits does not affect the result. \\
\indent
In the sum rule for the transversal polarization (27) the same interval
$0.7\,GeV^2 \le M^2\le 1.1\,GeV^2$ is investigated.  
The contribution of the dimension 4 operators gives not more
than 40\% of the total value, the contribution of the dimension 6 operators 
$-$ not more than 20\%. These contributions have opposite signs, at
$M^2=0.7\,GeV^2$ the last contribution is equal approximately to one half 
of the first one and reduces with increase of $M^2$.\\
\indent The increase of the value of the gluon
condensate by the factor 1.5 results in not more than 
 10\% rise of the value of the
second moment, the 
changing of the value of the quark condensate from 
$g^2\qq=0.23\times 10^{-2}\,GeV^6$ to
$g^2\qq=0.38\times 10^{-2}\,GeV^6$ increases ${\cal{M}}_\rho^T$
by $15\div 20\%$. \\ 
\indent A variation of $s_f$ in the interval $2.5-2.7\,GeV^2$ changes the 
value of the second moment ${\cal{M}}_\rho^T$ by $\pm 10\%$.\\
\indent One should evaluate also the uncertainties, which appear due to 
approximation procedure. In \cite{iof-3vertex} it was shown 
that nonlinear terms can be
safely neglected, when \\
${\cal{M}}_\rho^i\ll C_iM^2$, $i=L,T$. This condition
takes place in the case of the longitudinal polarization. In the transversal
polarization case ${\cal{M}}_\rho^T$ is only two times as large as $C_TM^2$ in 
the considered interval, and so we use chi-square technique to estimate 
deviations from linearity. Denoting the left hand side of equation (27) as
$R_T(M^2)$, we find:
$$\delta={1\over{{\cal{M}}_\rho^T}}\Big(\sum\limits_{j=1}^5(R_T(M_j^2)-
{\cal{M}}_\rho^T-C_TM_j^2)^2\Big)^{1/2}\approx 0.003\,,$$
where $M_1^2=0.7\,GeV^2,\;M_2^2=0.8\,GeV^2,...,M_5^2=1.1\,GeV^2$. 
Since $\delta\ll 1$, the nonlinear terms can be neglected
also in the case of the transversal polarization. \\
\indent Collecting all these uncertainties, one can find that the accuracy
of ${\cal{M}}_\rho^T$ is about 20\%.\\
\indent Our final results are:
$${\cal{M}}_\rho^L=0.84\pm 0.08\,,\eqno(31)$$
$${\cal{M}}_\rho^T=0.5\pm 0.1\,.\eqno(32)$$
\indent Now one can calculate the second moment ${\cal{M}}_\rho$ of
the non-polarized $\rho$-meson:
$${\cal{M}}_\rho=1/3{\cal{M}}_\rho^L+2/3{\cal{M}}_\rho^T\,.$$
This gives
$${\cal{M}}_\rho=0.60\pm 0.13\,.\eqno(33)$$
\indent Results (31), (32), (33) agree with the values obtained from the
lattice simulations\footnote{In order to compare the results, 
one should take into 
account a difference in choice of $\mu^2$.} \cite{sec-mom-lattice}.
It should be emphasized that the large difference between ${\cal{M}}_\rho^L$
and ${\cal{M}}_\rho^T$ (about 0.34, see (31), (32)) is close to those found
in  \cite{sec-mom-lattice} (about 0.29).\\
\indent  In \cite{iof-og-pion}, \cite{iof-og-rho} in the framework of QCD 
sum rules the valence quark structure functions were calculated at
the intermediate values of the Bjorken variable $x$. The sum rules had the
resembling structure, namely, in the case of the longitudinal polarization
the contributions of the quark and gluon condensates were quite small,
whereas in the case of the transversal polarization the dimension 4 operators
play an important role. \\
\indent The large value of ${\cal{M}}_\rho^L$
(31) lead to the conclusion that the gluon sea in the longitudinally
polarized $\rho$-meson is strongly suppressed ( it is less than 25\%, while
usually gluons carry about 50\% of the total momentum). 
So, it seems reasonable that the quark sea
in the longitudinally polarized $\rho$-meson is suppressed also. Therefore,
we can compare our result for ${\cal{M}}_\rho^L$ with the value 0.78,
 estimated in  \cite{iof-og-rho}. Let us remind that
in  \cite{iof-og-rho} the valence quark structure
functions was calculated in the interval $0.1<x<0.75$, whereas at small $x$
Regge asymptotic and at large $x$ quark counting rules were assumed. The 
fact that our result is quite close to the value of  \cite{iof-og-rho} 
verifies these assumptions. Thus, now the quark structure function can be
described in the whole region $0<x<1$.\\
\indent Unfortunately, the same analysis for the transversal polarization
can not be done, mainly because of the fact that in 
\cite{iof-og-rho} the valence quarks were considered, 
whereas in the present paper we can not separate valence and sea quarks. \\
\indent The second moment of the quark structure 
function of the pion ${\cal{M}}_\pi$ was calculated in \cite{bel-blok} 
and \cite{nishino}. However,
the contributions of the dimension 6 operators in these papers differ 
from each other. That is why we recalculate ${\cal{M}}_\pi$.\\
\indent ${\cal{M}}_\pi$ is related to the tensor $T_{\mu\nu}$
(1) in the similar way: 
$$\langle\pi|T_{\mu\nu}|\pi\rangle
=2{\cal{M}}_\pi p_{\mu}p_{\nu}\,,$$
$p_\mu$ is  the pion momentum.\\
\indent  Following \cite{bel-blok}, we consider the correlator of the 
axial currents $j_\mu=\overline{u}\gamma_\mu\gamma_5d$  
$$\Pi_{\mu\nu}(p)=i\int d^4x\,e^{ipx}\langle T(j_\mu(x)j^+_\nu(0)\rangle_S$$
and saturate its dispersion relation by the contributions of the pion and
continuum. We obtain the following sum rule:
$$\displaylines{{9\over{25}}(1-L^{50/81})+
{1\over{f_\pi^2}}L^{50/81}
\biggl({M^2\over{4\pi^2}}\Big(1-e^{-s_\pi/M^2}\Big)+{1\over{36}}
{\langle(\alpha_s/\pi)G^2\rangle\over{M^2}}+{34\over{81}}{g^2\qq\over{M^4}}
\bigg)=\hfill}$$
$$\displaylines{\hfill ={\cal{M}}_\pi(\mu^2)+C_\pi M^2\,.\;\;(34)}$$
In this equation $f_\pi=0.13\,GeV$, $s_\pi=0.9\,GeV$ is the continuum 
threshold for the pion and $C_\pi$ appears due to the non-diagonal 
transitions. From (30) we obtain:
$${\cal{M}}_\pi=0.6\pm 0.1\,.\eqno(35)$$
The contribution of the dimension 6 operators in (34) is very 
close\footnote{The small difference
can appear because of different ways of factorization.}
to the corresponding term in \cite{nishino} and strongly differs from
\cite{bel-blok}. 
The numerical value of ${\cal{M}}_\pi$ (35)
is in a good agreement with the experimental data \cite{m2-exp} (where the 
second moment of the valence quark structure function was obtained) and 
the results of lattice calculations \cite{sec-mom-lattice}.\\ 
\indent In  \cite{iof-og-rho} the value of the second moment of the valence
quark structure function of the pion was estimated: 0.44. It was found also 
that at the intermediate values
of the Bjorken variable the valence quark structure functions of the pion and
non-polarized $\rho$-meson are close to one another. In the present
paper we find that the second moments
of the (valence plus sea) quark structure functions of the pion and 
non-polarized $\rho$-meson coincide (see (33), (35)). 
Therefore, one can expect that their quark structure functions
are close in the whole interval of the Bjorken variable.\\
\indent It should be noted also that the second moments of the quark 
structure function of the pion and transversally polarized $\rho$-meson
are also quite close to one another (see (32) and (35)).  
\section{Conclusion}
$\,$
\indent We see that ${\cal{M}}_\rho^L>{\cal{M}}_\rho^T$
and the difference between them certainly
exceeds the accuracy of the results. So, polarization   
significantly influences the momentum distribution in the $\rho$-meson.\\
\indent The value of the second moment in the case of the longitudinal
polarization (31) allows one to determine the part of the momentum, carried
by the gluons: ${\cal{M}}_{\rho G}^L\sim 0.2$. Usually (in the pion or 
nucleon or transversally polarized $\rho$-meson) gluons carry about
one half of the total hadron momentum. It is for the first time that so 
small part of the momentum, carried by the gluons, was obtained. \\
\indent The estimations of the second moments in \cite{iof-og-rho} are in
consistent with our results, whereas the techniques of the calculations
have nothing in common. This fact confirms our present results and, in
general, points out to the selfconsistency of the QCD sum rule approach. \\
\indent The values of the second moments of quark structure functions of
the $\rho$-meson agree with the
results of lattice calculations. \\ \\ 
\indent The authors would like to thank B.L.\,Ioffe for statement of 
the problem and valuable discussions. \\
\indent The work is supported in part by grants CRDF RP2-2247, INTAS 2000
Project 587 and RFFI 00-02-17808.
\section{Appendix}
$\,$
\indent Here we consider the following vacuum expectation values of 
dimension 6 operators in the external field $S_{\rho\lambda}$:\\
$$\displaylines{
\langle\overline{q}_\alpha^a\hat G_{\mu\nu}D_\eta q_\beta^b\rangle_S\,,~~~~
\langle\overline{q}_\alpha^a\overleftarrow{D_\eta}\hat G_{\mu\nu}
q_\beta^b\rangle_S\,,~~~~
\langle\overline{q}_\alpha^a(D_\eta\hat G_{\mu\nu})q_\beta^b\rangle_S\,,
\hfill(A1)}$$
$$\displaylines{
\langle \overline q_\alpha^a D_{\{\sigma} D_\tau D_{\eta\}} q_\beta^b
\rangle_S\,,\hfill(A2)}$$
where $\hat G_{\mu\nu}=G_{\mu\nu}^nt^n$, the curly brackets mean 
symmetrization over the Lorenz indices and the round brackets in
$(D_\eta\hat G_{\mu\nu})$ indicate that derivative acts only on 
$\hat G_{\mu\nu}$.\\
\indent 
Let us start with the derivation of some useful for the further  analysis
relations. \\
\indent Using C-parity, one can easily see that\footnote{Hereafter we 
omit spinor and color indices, if the summation over them is assumed.}
$$\langle\overline q^a\gamma_\sigma\hat G_{\mu\nu}D_\eta q^b\rangle_S=
{1\over 2}\Big(\langle\overline q^a\gamma_\sigma
\hat G_{\mu\nu}D_\eta q^b\rangle_S+
\langle\overline q^a\overleftarrow{D_\eta}\gamma_\sigma
\hat G_{\mu\nu}q^b\rangle_S\Big)\eqno(A3)$$
and
$$\langle\overline q^a\gamma_5\gamma_\sigma\hat G_{\mu\nu}D_\eta q^b\rangle_S=
{1\over 2}\Big(\langle\overline q^a\gamma_5\gamma_\sigma\hat G_{\mu\nu}
D_\eta q^b
\rangle_S-
\langle\overline q^a\overleftarrow{D_\eta}\gamma_5\gamma_\sigma
\hat G_{\mu\nu}q^b\rangle_S\Big)\,.\eqno(A4)$$
From the other side, with the accuracy up to full derivative one can write:
$$\langle\overline q^a\gamma_\sigma\hat G_{\mu\nu}D_\eta q^b\rangle_S=
-\langle\overline q^a\overleftarrow{D_\eta}\gamma_\sigma
\hat G_{\mu\nu}q^b\rangle_S-
\langle\overline q^a\gamma_\sigma(D_\eta\hat G_{\mu\nu})q^b\rangle_S\,,
\eqno(A5)$$
$$\langle\overline q^a\gamma_5\gamma_\sigma\hat G_{\mu\nu}D_\eta q^b\rangle_S=
-\langle\overline q^a\overleftarrow{D_\eta}\gamma_5\gamma_\sigma
\hat G_{\mu\nu}q^b\rangle_S-\langle\overline q^a\gamma_5
\gamma_\sigma(D_\eta\hat G_{\mu\nu})q^b\rangle_S\,.\eqno(A6)$$
Comparing $(A4)$ and $(A6)$, $(A5)$ and $(A3)$, we see that
$$\langle\overline q^a\gamma_\sigma\hat G_{\mu\nu}D_\eta q^b\rangle_S=
\langle\overline q^a\overleftarrow{D_\eta}\gamma_\sigma
\hat G_{\mu\nu}q^b\rangle_S=-{1\over 2}
\langle\overline q^a\gamma_\sigma(D_\eta\hat G_{\mu\nu})q^b\rangle_S\,,$$
$$\langle\overline q^a\gamma_5\gamma_\sigma\hat G_{\mu\nu}D_\eta q^b\rangle_S=
-\langle\overline q^a\overleftarrow{D_\eta}\gamma_5\gamma_\sigma
\hat G_{\mu\nu}q^b\rangle_S\,,\eqno(A7)$$
$$\langle\overline q^a\gamma_5\gamma_\sigma(D_\eta\hat G_{\mu\nu})q^b\rangle_S
=0\,.$$
\indent Now let us write down the general tensor structure of 
vacuum expectation values $(A1)$.
$$\displaylines{\langle \overline{q}_\alpha^a G_{\eta\sigma}^nD_\tau 
q_\beta^b\rangle_S=
(t^n)^{ba}\Big(d_1(S_{\tau\eta}\gamma_\sigma
-S_{\tau\sigma}\gamma_\eta)+
d_2(g_{\tau\eta}S_{\sigma\xi}-g_{\tau\sigma}S_{\eta\xi})
\gamma_\xi+\hfill}$$
$$\displaylines{\hfill 
+id_3\epsilon_{\eta\sigma\xi\chi}S_{\tau\xi}\gamma_\chi \gamma_5+
id_4\epsilon_{\tau\eta\sigma\xi}S_{\xi\chi}\gamma_\chi \gamma_5
\Big)_{\beta\alpha}\;,}$$

$$\displaylines{\langle \overline{q}_\alpha^a
\overleftarrow{D_\tau}G_{\eta\sigma}^n
q_\beta^b\rangle_S=
(t^n)^{ba}\Big(e_1(S_{\tau\eta}\gamma_\sigma-S_{\tau\sigma}\gamma_\eta)+
e_2(g_{\tau\eta}S_{\sigma\xi}-g_{\tau\sigma}S_{\eta\xi})
\gamma_\xi+\hfill}$$
$$\displaylines{\hfill 
+ie_3\epsilon_{\eta\sigma\xi\chi}S_{\tau\xi}\gamma_\chi \gamma_5+
ie_4\epsilon_{\tau\eta\sigma\xi}S_{\xi\chi}\gamma_\chi \gamma_5
\Big)_{\beta\alpha}\;,~~~~(A8)}$$

$$\displaylines{\langle \overline{q}_\alpha^a
(D_\tau G_{\eta\sigma}^n)q_\beta^b\rangle_S=
(t^n)^{ba}\Big(f_1(S_{\tau\eta}\gamma_\sigma
-S_{\tau\sigma}\gamma_\eta)+
f_2(g_{\tau\eta}S_{\sigma\xi}-g_{\tau\sigma}S_{\eta\xi})\gamma_\xi+\hfill}$$
$$\displaylines{\hfill 
+if_3\epsilon_{\eta\sigma\xi\chi}S_{\tau\xi}\gamma_\chi \gamma_5+
if_4\epsilon_{\tau\eta\sigma\xi}S_{\xi\chi}\gamma_\chi \gamma_5
\Big)_{\beta\alpha}\;,}$$
where $d_1,...,d_4$, $e_1,...,e_4$, $f_1,...,f_4$ are unknown constants.
All other possible structures can be expressed in terms of these one.
Multiplying these equations by $(\gamma_\sigma)_{\alpha\beta}$,
$g_{\tau\eta}(\gamma_\rho)_{\alpha\beta}$,
$\epsilon_{\eta\sigma\mu\nu}(\gamma_5 \gamma_\rho)_{\alpha\beta}$
and so on and using $(A7)$, we find after simple algebra:
$$d_1=e_1=-1/2f_1\,,~~~~d_2=e_2=-1/2f_2\,,~~~~f_3=f_4=f_5=0\,,$$
$$d_3=-e_3\,,~~~~d_4=-e_4\,.$$
\indent One can find also the relation between $d_1$ and $d_2$. Multiplying 
the third of equations $(A8)$ by $(\gamma_\chi)_{\alpha\beta}g_{\tau\eta}
(t^n)^{ab}$,
we find:
$$\langle \overline q_\alpha^a(D_\tau G_{\eta\sigma}^n)q_\beta^b\rangle_S
(\gamma_\chi)_{\alpha\beta}g_{\tau\eta}(t^n)^{ab}
=-g\langle(\overline q\gamma_\chi t^n q)
(\overline q\gamma_\sigma t^n q)\rangle_S\,.$$
Using factorization hypothesis, $\langle(\overline q\gamma_\chi t^n q)
(\overline q\gamma_\sigma t^n q)\rangle_S$ can be expressed in terms of 
$\langle\overline q_\alpha^a q_\beta^b\rangle_S$, which equals to zero, since 
the proportional to the external field $S_{\mu\nu}$ tensor structure 
does not exist. We obtain:
$$d_1=3d_2\,.$$
Thus, all vacuum expectation values $(A8)$ are expressed in terms of three
unknown constants (for example, $d_1,\,d_3,\,d_4$).\\
\indent Therefore, we obtain for the vacuum expectation values $(A8)$:
$$\displaylines{\langle \overline{q}_\alpha^a G_{\eta\sigma}^nD_\tau 
q_\beta^b\rangle_S=
(t^n)^{ba}\Big(d_1(S_{\tau\eta}\gamma_\sigma
-S_{\tau\sigma}\gamma_\eta+
{1/3}(g_{\tau\eta}S_{\sigma\xi}-g_{\tau\sigma}S_{\eta\xi})
\gamma_\xi)+\hfill}$$
$$\displaylines{\hfill 
+id_3\epsilon_{\eta\sigma\xi\chi}S_{\tau\xi}\gamma_\chi \gamma_5+
id_4\epsilon_{\tau\eta\sigma\xi}S_{\xi\chi}\gamma_\chi \gamma_5
\Big)_{\beta\alpha}\;,}$$

$$\displaylines{\langle \overline{q}_\alpha^a
\overleftarrow{D_\tau}G_{\eta\sigma}^n
q_\beta^b\rangle_S=
(t^n)^{ba}\Big(d_1(S_{\tau\eta}\gamma_\sigma-S_{\tau\sigma}\gamma_\eta+
{1/3}(g_{\tau\eta}S_{\sigma\xi}-g_{\tau\sigma}S_{\eta\xi})
\gamma_\xi)-\hfill}$$
$$\displaylines{\hfill 
-id_3\epsilon_{\eta\sigma\xi\chi}S_{\tau\xi}\gamma_\chi \gamma_5-
id_4\epsilon_{\tau\eta\sigma\xi}S_{\xi\chi}\gamma_\chi \gamma_5
\Big)_{\beta\alpha}\;,~~~~(A9)}$$

$$\langle \overline{q}_\alpha^a
(D_\tau G_{\eta\sigma}^n)q_\beta^b\rangle_S=
-2d_1(t^n)^{ba}\Big(S_{\tau\eta}\gamma_\sigma
-S_{\tau\sigma}\gamma_\eta+
{1/3}
(g_{\tau\eta}S_{\sigma\xi}-g_{\tau\sigma}S_{\eta\xi})\gamma_\xi
\Big)_{\beta\alpha}\;.$$
For compliteness we express $d_1,\,d_3,\,d_4$ in terms of the unknown
 vacuum expectation values:
$$d_1S_{\tau\mu}={3\over{2^7}}\langle\overline q\gamma_\sigma 
\hat G_{\mu\sigma}D_\tau q\rangle_S\,,$$
$$d_3S_{\tau\mu}={i\over{2^8}}\epsilon_{\eta\sigma\mu\rho}
\Big(3\langle\overline q\gamma_5\gamma_\rho 
\hat G_{\eta\sigma}D_\tau q\rangle_S-
\langle\overline q\gamma_5\gamma_\tau 
\hat G_{\eta\sigma}D_\rho q\rangle_S\Big)\,,$$
$$d_4S_{\tau\mu}={i\over{2^8}}\epsilon_{\eta\sigma\mu\rho}
\Big(\langle\overline q\gamma_5\gamma_\rho 
\hat G_{\eta\sigma}D_\tau q\rangle_S-
 3\langle\overline q\gamma_5\gamma_\tau 
\hat G_{\eta\sigma}D_\rho q\rangle_S\Big)\,.$$
\indent Multiplying now the first of equations $(A9)$ by 
$(\gamma_\tau)_{\gamma\beta}(t^n)^{ab}$ and using equation of motion
$$\hat{D}q-S_{\mu\nu}D_\mu\gamma_\nu q=0\,,\eqno(A10)$$
we obtain additional relation:
$${4\over 3}d_1-d_3+d_4=-{g\qq\over{3^32^4}}\,.\eqno(A11)$$
\indent Now we consider the vacuum expectation value $(A2)$
$$\langle\overline{q}_\alpha^a(D_{\{\sigma} D_\tau D_{\eta\}}q_\beta^b
\rangle_S=(c_1S_{\{\sigma\tau}\gamma_{\eta\}}+
c_2g_{\{\sigma\tau}S_{\eta\}\chi}\gamma_\chi)_{\beta\alpha}\delta^{ab}\,.$$
Multiplying this relation by $(\gamma_\eta)_{\gamma\beta}$ and taking into
account $(A9)$, $(A10)$ and the identity 
$[D_\mu D_\nu]=-igt^nG_{\mu\nu}^n$, we find
$$c_1=-3k(1+\epsilon_1)\,,$$
$$c_2=3k(1+\epsilon_2)\,,\eqno (A12)$$
where
$$k=-{ig^2\qq\over{3^52^4}}\,,$$
$$\epsilon_1={ig\over{3k}}\Big(-d_1+{d_3\over 3}\Big)\,,\eqno(A13)$$
$$\epsilon_2={ig\over{3k}}\Big({7\over 9}d_1+d_3)\,.$$
In $(A12)$ the first term appears due to 
$\langle\overline q_\alpha^a D_\mu D_\nu \hat Dq_\beta^b\rangle$, 
which can be calculated
by use of the equation of motion (the factorizable vacuum expectation value),
while $\epsilon_1$, $\epsilon_2$, expressed in terms of 
$d_1,\,d_3,\,d_4$, correspond to contributions of the nonfactorizable
 vacuum expectation values (i.e. to contributions of the
 vacuum expectation values, which can not be calculated in such a way).\\
\indent Thus, vacuum expectation values $(A1)$, $(A2)$ are expressed 
in terms of three unknown parameters, related by $(A11)$.

\newpage
\begin{figure}
\epsfxsize=6.0cm
\epsfbox{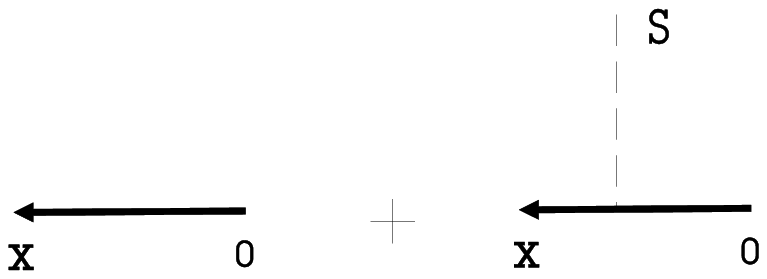}
\caption{Diagrams for the quark propagator (solid line) in the external tensor 
field (dashed line).}
\end{figure}
\begin{figure}
\epsfxsize=6.0cm
\epsfbox{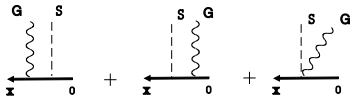}
\caption{Diagrams for the quark propagator (solid line) in the external tensor 
field (dashed line) and soft gluon field (wave line).}
\end{figure}

\newpage
\begin{figure}
\epsfxsize=6.0cm
\epsfbox{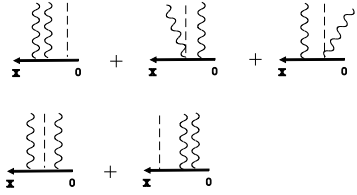}
\caption{Diagrams for the quark propagator in the external tensor 
field and soft gluon field. All notations are as in fig.2.}
\end{figure}
\begin{figure}
\epsfxsize=6.0cm
\epsfbox{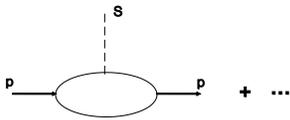}
\caption{Loop diagrams in the external tensor field. $p$ represents the
momentum of the current. Other notations are as in fig.1.}
\end{figure}

\newpage
\begin{figure}
\epsfxsize=6.0cm
\epsfbox{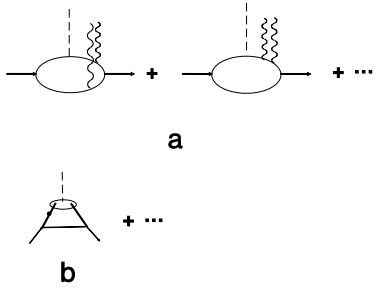}
\caption{Diagrams of dimension 4 operators. Black circle on the
quark line means derivative. Dots stand for permutations. Other 
notations are as in fig.2.}
\end{figure}

\begin{figure}
\epsfxsize=6.0cm
\epsfbox{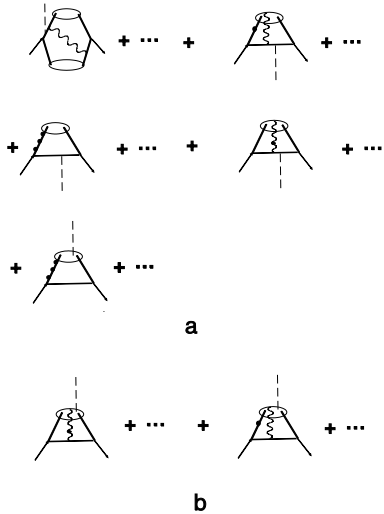}
\caption{Diagrams of dimension 6 operators. Black circles on the
quark or gluon lines mean derivatives. All other 
notations are as in the previous figures.}
\end{figure}

\newpage
\begin{figure}
\epsfxsize=6.0cm
\epsfbox{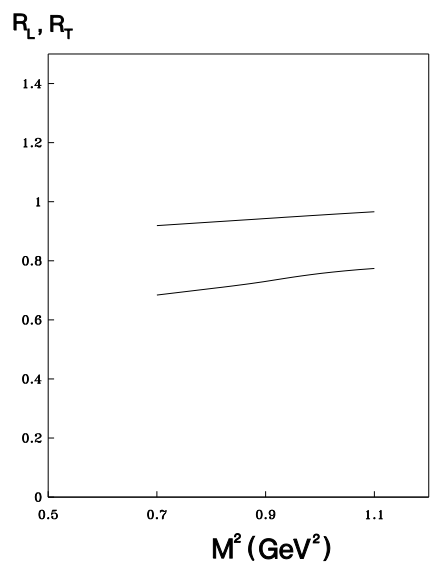}
\caption{The left hand sides of equations (26) $R_L$ (the thin curve), 
(27) $R_T$ (the thick curve) as functions of squared Borel mass $M^2$.}
\end{figure}
\end{document}